\documentclass[aps, prapplied,reprint,groupedaddress, longbibliography]{revtex4-1}
\usepackage{amsmath,bm}
\usepackage{amsmath}
\usepackage[colorlinks=true]{hyperref}
\usepackage{amssymb} 
\usepackage{mathrsfs}
\usepackage{graphicx}
\usepackage{epstopdf}
\usepackage{graphicx}
\usepackage{dcolumn}
\usepackage{bm}
\usepackage{float}
\usepackage{soul}
\usepackage{color}
\usepackage{siunitx}
\usepackage{booktabs}

\usepackage[dvipsnames]{xcolor}

\makeatletter
\newcommand*{\rom}[1]{\expandafter\@slowromancap\romannumeral #1@}
\makeatother

\graphicspath{{figures/}}

\begin{document}


\title{Improving sound absorption through nonlinear active electroacoustic resonators}


\author{Xinxin Guo}
\email{xinxin.guo@epfl.ch}
\affiliation{Laboratoire de Traitement des Signaux LTS2, Ecole Polytechnique F\'ed\'erale de Lausanne, 1015 Lausanne, Switzerland}
\author{Romain Fleury}
\email{romain.fleury@epfl.ch}
\affiliation{Laboratory of Wave Engineering, Ecole Polytechnique F\'ed\'erale de Lausanne, 1015 Lausanne, Switzerland}
\author{Herv\'e Lissek}
\email{herve.lissek@epfl.ch}
\affiliation{Laboratoire de Traitement des Signaux LTS2, Ecole Polytechnique F\'ed\'erale de Lausanne, 1015 Lausanne, Switzerland}


\date{\today}

\begin{abstract}
Absorbing airborne noise at frequencies below 300 Hz is a particularly vexing problem due to the absence of natural sound absorbing materials at these frequencies. The prevailing solution for low-frequency sound absorption is the use of passive narrow-band resonators, whose absorption level and bandwidth can be further enhanced using nonlinear effects. However, these effects are typically triggered at high intensity levels, without much control over the form of the nonlinear absorption mechanism. In this study, we propose, implement, and experimentally demonstrate a nonlinear active control framework on an electroacoustic resonator prototype, allowing for unprecedented control over the form of non-linearity, and arbitrarily low sound intensity thresholds. More specifically, the proposed architecture combines a linear feedforward control on the front pressure through a first microphone located at the front face of the loudspeaker, and a nonlinear feedback on the membrane displacement estimated through the measurement of the pressure inside the back cavity with a second microphone located in the enclosure. It is experimentally shown that even at a weak excitation level, it is possible to observe and control the nonlinear behaviour of the system. Taking the cubic nonlinearity as an example, we demonstrate numerically and experimentally that in the low frequency range ($[\SI{50}{Hz}, \SI{500}{Hz}]$), the nonlinear control law allows improving the sound absorption performance, i.e. enlarging the bandwidth of optimal sound absorption while increasing the maximal absorption coefficient value, and producing only a negligible amount of nonlinear distortion. The reported experimental methodology can be extended to implement various types of hybrid linear and/or nonlinear controls, thus opening new avenues for managing wave nonlinearity and achieving non-trivial wave phenomena.
\end{abstract}



\maketitle

\section{Introduction} \label{intro}
Broadband sound absorption, especially in low frequencies,
still remains a challenge in both scientific research and engineering
practice. Conventional sound absorbing materials, such as porous and fibrous media \cite{Biot_1956, Allard_2009}, are not efficient for achieving effective absorption at low frequencies with thin layers, due to the causal nature of the acoustic response that dictates a sum rule relating the absorption spectrum and the sample thickness \cite{Landau_1984, fano_theoretical_1950, causality_IEEE, broadband_scattering_JASA, Psheng_horizons}. Moreover, for any passive, linear and time-invariant system, the bandwidth and the absorption efficiency are mutually constrained, consistent with the 'Bode-Fano criterion' \cite{Bode_1945, Pozar_1998, absorp_constraint_2009, Bound_cloaking_2016}. Bypassing such inherent bounds by revoking their underlying assumptions has allowed the design of wideband matching devices \cite{Active_Alu_2013,Active_IEEE,BodeFano_PRL}.

Amongst these assumptions, passivity was first considered. It can be violated by actively controlling the acoustic features of systems \cite{Active_Alu_2013,Active_IEEE}. Such control, applied to Electroacoustic Resonators (ERs) to enable impedance adjustment, allows broadening the sound absorption, especially in the low frequency range \cite{etienne_IEEE, herve_2011,etienne_MDOF, etienne_acta, boulandet_2010}. A wide range of achievable acoustic impedances can be provided by the concept, including the synthesis of narrow-band single-degree of freedom (SDOF) resonators \cite{boulandet_2010}, resonances with multiple degrees of freedom \cite{etienne_MDOF,etienne_acta}, or a high degree of reconfigurability \cite{Romain_prapplied_2019}. Such tunability is key in many applications, such as room mode damping \cite{rivet_aes_2016,etienne2016} or aircraft engine tonal noise reduction \cite{boulandet_jsv_2018}. Recently, active control has also received a surge of interest as a tool for designing Acoustic Metamaterials (AMMs) that overcome the restrictions imposed by passive AMMs \cite{romain_review2019,active_membrane_PRB,piezo_Adv_material,cummer_review_2016}, thereby expanding the reach of metamaterial science to a wealth of nontrivial acoustic phenomena such as PT-symmetry scattering \cite{romain_invisible_2015,PT_symmetry_PRL,PT_symmetry_PRX,PhysRevApplied2017}, wavefront shaping \cite{Active_metasurface_PSheng, wave_shaping_PRL,lissek_JAP_2018} and non-Hermitian wave control \cite{etienne_constant_pressure_2018,constant_pressure_NC}. A notable technique for active impedance control uses a conventional loudspeaker, whose acoustic impedance can be modified either by shunting its electric terminals with an engineered electric load \cite{etienne_IEEE,boulandet_2010, etienne2016}, or by feeding back a current/voltage proportional to a combination of sensed acoustic quantities \cite{herve_2011, control_1997, etienne2016}.

In the field of Active Electroacoustic Resonators (AERs), most of the previous studies have been carried out under the assumption that the involved acoustic parameters fluctuations were small enough to ensure that they remain linear at low frequencies. Nevertheless, nonlinear resonators exhibit also interesting performances that contribute to a variety of wave phenomena. For instance, a primary linear resonator coupled with a purely nonlinear resonator, known as Nonlinear Energy Sink (NES) \cite{gendelman_2000,Gendelman2001,gendelman_JSV,Gourdon_2007} enables vibration extinction of the linear system, a phenomenon called energy pumping or targeted energy transfer \cite{gendelman_NLDynamics, Mattei_2010, Mattei_2011, Cote_2014,Bryk_JSV}. Typical nonlinear effects such as higher harmonic generation have been demonstrated and investigated in metamaterials made of nonlinear resonators \cite{xinxin_JAP, xinxin_PRE,NL_metamaterials_JASA}. However, the aforementioned systems usually do not allow tunable nonlinear behavior, especially at low intensities, and are typically associated with large intensity thresholds. Unlike for electromagnetic signals for which nonlinearity has been exploited and incorporated with active control \cite{romain_NLinsulator_2019,NL_PT_perspective}, the possibility to create acoustic resonators with tunable nonlinear response \cite{NL_PT_perspective,non-reciprocal_NL_2014} has been left largely unexplored, except in a recent numerical study by D. Bitar et al. \cite{diala_JSV}.

In this paper, we establish both numerically and experimentally a nonlinear control methodology that enables achieving a controllable nonlinear SDOF AER, that exhibits nonlinear effects even at weak excitation levels. This is obtained through a current-driven feedback control framework applied on a closed-box electrodynamic loudspeaker. Focusing on the sound absorption performance, we use our findings and determine proper control laws that allow improving sound absorption, while producing only a negligible amount of nonlinear distortion.

The paper is organized as follows. Based on the known linear theory of active impedance control on the ER, a nonlinear control strategy is firstly introduced in Section \ref{SDOF_NL}, together with the definition of a relevant absorption performance metric. The prototype and the corresponding experimental set up are then described in Section \ref{set up}. Thereafter, the absorption performance of the achieved nonlinear AER prototype is examined. Two different types of control laws are considered; a purely nonlinear control law (Section \ref{NL control}) and hybrid control laws that combine different linear settings with the proposed nonlinear one (Section \ref{hybrid control}). Simulation through a time-domain integration method is also implemented in order to verify and validate the experimental results.   

\section{Nonlinear single degree-of-freedom electroacoustic resonator} \label{SDOF_NL}
\subsection{Description and working principle}\label{formulation}
\begin{figure}
	\includegraphics[width=8.5cm]{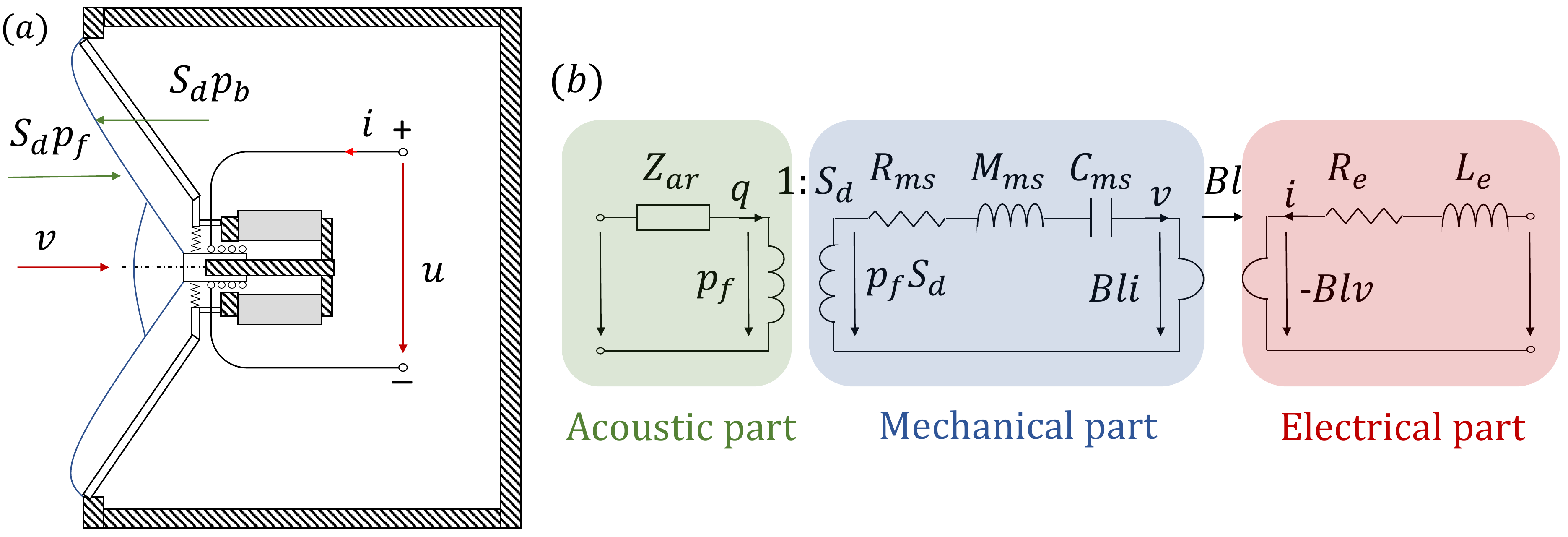}
	\centering
	\caption{\label{fig1} Schematic representation (a) and circuit analogy (b) of the considered closed-box electrodynamic loudspeaker system (source: Rivet, 2016\cite{etienne2016}).}
\end{figure}

In the low frequency range and under weak excitation, an electrodynamic loudspeaker behaves as a linear SDOF ER. The mechanical part of the loudspeaker can be simply modeled as a conventional mass-spring-damper system, where the moving diaphragm of mass $M_{ms}$ is attached through an elastic suspension of mechanical compliance $C_{ms}$, and the global losses are accounted for in the mechanical resistance $R_{ms}$. In the present work, we consider a loudspeaker mounted in an enclosure of volume $V_b$. Fig.~\ref{fig1} illustrates the schematic representation and the circuit analogy of the closed-box loudspeaker. Denoting $S_d$ the effective area of the loudspeaker diaphragm and $Bl$ the force factor of the moving-coil transducer, the Newton's second law, applied to the loudspeaker diaphragm, reads:
\begin{align}
  M_{ms} \frac{d v(t)}{d t} = & S_{d}(p_f(t)-p_b(t))-R_{ms}v(t)-\frac{1}{C_{ms}}\int v(t)dt \nonumber  \\
  &-Bli(t),
\label{eq1}  
\end{align}
where $p_f(t)$ and $p_b (t)$ designate the acoustic pressures applied respectively to the front and the rear faces of the membrane, whereas $v(t)$ and $i(t)$ represent the acoustic velocity of the diaphragm and the current circulating in the moving coil, respectively.

At low frequencies, the sound pressure inside the cavity of volume $V_b$ is assumed uniform, yielding a linear relation between the rear pressure $p_b(t)$ and the displacement of diaphragm $\xi(t)=\int v(t)dt$, i.e.,
\begin{equation}
p_b(t) \cong \frac{S_d}{C_{ab}}\xi(t),
\label{eq2}
\end{equation}
with $C_{ab}=V_b/(\rho c^2)$ representing the acoustic compliance of the enclosure, where $\rho$ and $c$ denote the air mass density and the associated speed of sound. Introducing the overall mechanical compliance $C_{mc}=C_{ms}C_{ab}/(S_d^2C_{ms}+C_{ab})$ accounting for the fluid compressibility on the rear face of diaphragm, Eq.~\eqref{eq1} can be rewritten as:  
\begin{equation}
M_{ms} \frac{d^2 \xi(t)}{dt^2} =p_f(t)S_{d}-R_{ms}\frac{d \xi(t)}{d t}-\frac{1}{C_{mc}}\xi(t)-Bli(t).
\label{eq3}
\end{equation}

In the open circuit configuration, i.e., $i=0$, the frequency response of the considered ER is characterized by the specific acoustic impedance $Z_{as}$ defined in the linear regime by
\begin{equation}
Z_{as}(j\omega)=\frac{P_f(j\omega)}{V(j\omega)}=j\omega M_{as}+R_{as}+\frac{1}{j\omega C_{ac}},
\label{eq4}
\end{equation}
where $M_{as}= M_{ms}/S_d$, $R_{as}= R_{ms}/S_d$ and $C_{ac}= C_{mc}S_d$ are the equivalent acoustic parameters. The uppercase symbols $P_f$ and $V$ designate the frequency responses of the considered acoustic quantities (front pressure and velocity) to distinguish with their denotations in the time domain (represented by lowercase symbols).

When the loudspeaker is driven with a given electrical current, the added term due to $i(t) \neq 0$ leads to an impedance response different from $Z_{as}$. Active impedance control is typically based on controlling the current that circulates through the loudspeaker coil. This type of control has proven to be more stable compared to others such as voltage control \cite{etienne2016}, as it offers the opportunity to tune the acoustic impedance and the absorption performance of the acoustic resonator without having to model its electrical part.

\subsection{Acoustic impedance control of a linear AER} \label{Active control}
Before introducing the nonlinear control strategy, a case of linear control is first presented. To this end, one defines a target specific acoustic impedance $Z_{st}$ that the linear AER, once controlled, is expected to present. We assume here that it takes the form of a SDOF resonator, similar to the passive impedance of Eq.~\eqref{eq4}:
\begin{equation}
Z_{st}(j\omega)=j\omega \mu_1 M_{as}+R_{st}+\frac{\mu_2}{j\omega C_{ac}},
\label{eq5}
\end{equation}
where $\mu_1$, $\mu_2$ and $R_{st}$ are design parameters corresponding to the desired mass, compliance and resistance of the controlled ER, respectively. Such target impedance parameters are used to adjust the frequency response of resonator (resonance frequency, quality and damping factor). Regarding the absorption performance, it is easy to show that the maximum absorption appears at the frequency \cite{etienne2016}:
\begin{equation}
f_{st}=f_{s}\sqrt{\frac{\mu_2}{\mu_1}},
\label{eq5_1}
\end{equation}
where $f_s=(2\pi \sqrt{M_{as}C_{ac}})^{-1}$ is the natural resonance frequency of the ER. Thus, by adjusting the ratio $\mu_2/\mu_1$, the frequency of maximum absorption can be tuned (note that it can also be left unchanged to $f_s$). Additionally, perfect absorption can be achieved as well at prescribed frequency $f_{st}$ if the target resistance $R_{st}$ reaches the specific acoustic impedance of air, i.e., $R_{st}=Z_c=\rho c$. 

The objective of active impedance control is to identify the controller transfer function enabling the desired conversion from the input pressure $p_f$, that is sensed using a microphone, to the output current $i$ that is sent back to the loudspeaker, in order to achieve the target impedance $Z_{st}$ on the ER. In the considered linear regime, the transfer function $\Phi(s)$ can be derived from Eq.~\eqref{eq3} and Eq.~\eqref{eq5} in the Laplace domain (with variable $s$) as:
 \begin{equation}
\Phi(s) = \frac{I_L(s)}{P_f(s)}=\frac{Z_{st}(s)S_d-Z_{as}(s)S_d}{BlZ_{st}(s)},
\label{eq6}
\end{equation}
where the symbol $I_L$ denotes the Laplace transform of the current in the linear configuration.

Through the control of the current $I_L$ delivered to the loudspeaker terminals as a function $\Phi(s)$ of the input front pressure $P_f$, the impedance and the absorption properties of the resonator can therefore be tuned. This was previously demonstrated over a quite wide frequency range, depending on the control parameters ($\mu_1,\mu_2,R_{st}$) \cite{etienne_IEEE}.

\subsection{Nonlinear control of the ER}
As previously seen in Eq.~\eqref{eq2}, the rear pressure $p_b$ is proportional to the displacement of the loudspeaker diaphragm in the low frequency range. This provides the opportunity to define a current $i_{NL}$ as a function of a nonlinear transformation of the rear pressure. In the present work, we propose to add a nonlinear part to the control law by driving an additional current $i_{NL}$, defined as a nonlinear cubic transformation of the rear pressure $p_b$
\begin{equation}
i_{NL}(t)=G_{ui}\times\beta_{NL}\times(G_{mic} p_b(t))^3 \propto \xi^3(t),
\label{eq7}
\end{equation}
where $\beta_{NL}$ denotes the tunable nonlinear parameter, while $G_{mic}$ and $G_{ui}$ are the sensitivity of the microphone and the gain that converts the voltage into current, respectively.

Then, such current $i_{NL}$ will contribute to adding a nonlinear component to the stiffness (inverse of compliance) of the resonator, that would be fully adjustable and potentially much larger than what is possible with passive mechanical elements. Indeed, for an intrinsically nonlinear mechanical system, a relatively strong excitation is always required to trigger nonlinear effects. Here instead, by simply increasing the nonlinear parameter $\beta_{NL}$, the proposed nonlinear control methodology facilitates the emergence of nonlinear phenomena without requiring large excitation levels. Implementation and further analysis of the designed nonlinear AER are presented in the Section \ref{NL control} and the Section \ref{hybrid control}, by considering either a pure nonlinear control defined with $i=i_{NL}$, as well as a hybrid control with $i=i_{L}+i_{NL}$.

\subsection{Sound absorption coefficient metric for linear and nonlinear AERs}\label{Sec:alpha}

We are interested in the sound absorption performance of the controlled AERs. The chosen metric of interest is the sound absorption coefficient, which can be determined by sensing both the front pressure and the membrane axial velocity. 

In the linear regime, such two quantities, measured in the time domain and processed in the frequency domain, allow the effective specific acoustic impedance of the diaphragm $Z(j\omega) =P_f(j\omega)/V(j\omega)$ to be easily extracted under a sweep sine excitation over the frequency range of interest. Then, the sound absorption coefficient $\alpha_L(j\omega)$ can be obtained in a straightforward manner through the conventional relationship:
\begin{equation}
\alpha_L(j\omega)=1-\mid \dfrac{Z(j\omega)-Z_c}{Z(j\omega)+Z_c} \mid^2,
\label{eq9}
\end{equation}
where $Z_c=\rho c$ denotes the specific acoustic impedance of air.

However, for a nonlinear system, the energy transfer from fundamental frequency $\omega$ to higher harmonics ($n\omega$ with $n \geq 2$) should be additionally taken into account, leading to a generalized definition of the absorption coefficient as:

\begin{equation}
\alpha_{NL}=1-\sum_{n=1}^{n=N}\mid R_n\mid^2=\alpha_L-\sum_{n=2}^{n=N}\mid R_n\mid^2,
\label{eq10}
\end{equation}
where $R_n$ represents the complex pressure amplitude of the generated n-th harmonic normalized by that of the fundamental incoming wave.

\section{experimental set up}\label{set up} 
In our experiment, a commercially-available electrodynamic loudspeaker (Monacor SPX-30M), mounted with an enclosure having lateral surface of $\SI{12}{cm}\times\SI{12}{cm}$ and with thickness of $\SI{6.8}{cm}$, is employed as the experimental ER prototype. The overall closed-box ER presents a resonance frequency around \SI{200}{Hz}, corresponding to a wavelength of $\SI{1.7}{m}$ which is $25$ times larger than the cavity dimensions, confirming the sub-wavelength nature of the absorber. Moreover, notice that the definition of the linear control law requires knowing the mechanical parameters $M_{ms}$, $R_{ms}$ and $C_{mc}$ as well as the force factor $Bl$ of the considered ER (see Eq.~\eqref{eq5} and Eq.~\eqref{eq6}). These parameters are determined from two calibration measurements of the acoustic impedance, the first obtained with the ER in open circuit and the second in short circuit case, as presented in Ref.~\cite{etienne2016}. The extracted loudspeaker parameters, as well as the estimated effective area $S_d$ of the loudspeaker diaphragm, are summarized in Table.~\ref{table0}

\begin{table}[htbp]
\caption{\label{table0} Estimated Thiele-Small parameters of the closed-box Monacor SPX-30M lousdpeaker.}
\begin{center}
\begin{ruledtabular}
\begin{tabular}{lccccc}
\textbf{Parameter}  & \multicolumn{1}{c}{$M_{ms}$} & \multicolumn{1}{c}{$R_{ms}$} & \multicolumn{1}{c}{$C_{mc}$} & \multicolumn{1}{c}{$B\ell$} & \multicolumn{1}{c}{$S_d$} \\
\hline
\rule{0pt}{12pt} \textbf{Unit} &  \multicolumn{1}{c}{g} &  \multicolumn{1}{c}{N.s.m$^{-1}$} &  \multicolumn{1}{c}{mm.N$^{-1}$} &  \multicolumn{1}{c}{N.A$^{-1}$} &  \multicolumn{1}{c}{cm$^2$}\\
\hline
\rule{0pt}{12pt} \textbf{Value} &  \multicolumn{1}{c}{2.7} &  \multicolumn{1}{c}{0.4516} &  \multicolumn{1}{c}{0.2185} &  \multicolumn{1}{c}{3.3877} &  \multicolumn{1}{c}{32}\\
\end{tabular}
\end{ruledtabular}
\end{center}
\end{table}

\begin{figure}
	\includegraphics[width=8.5cm]{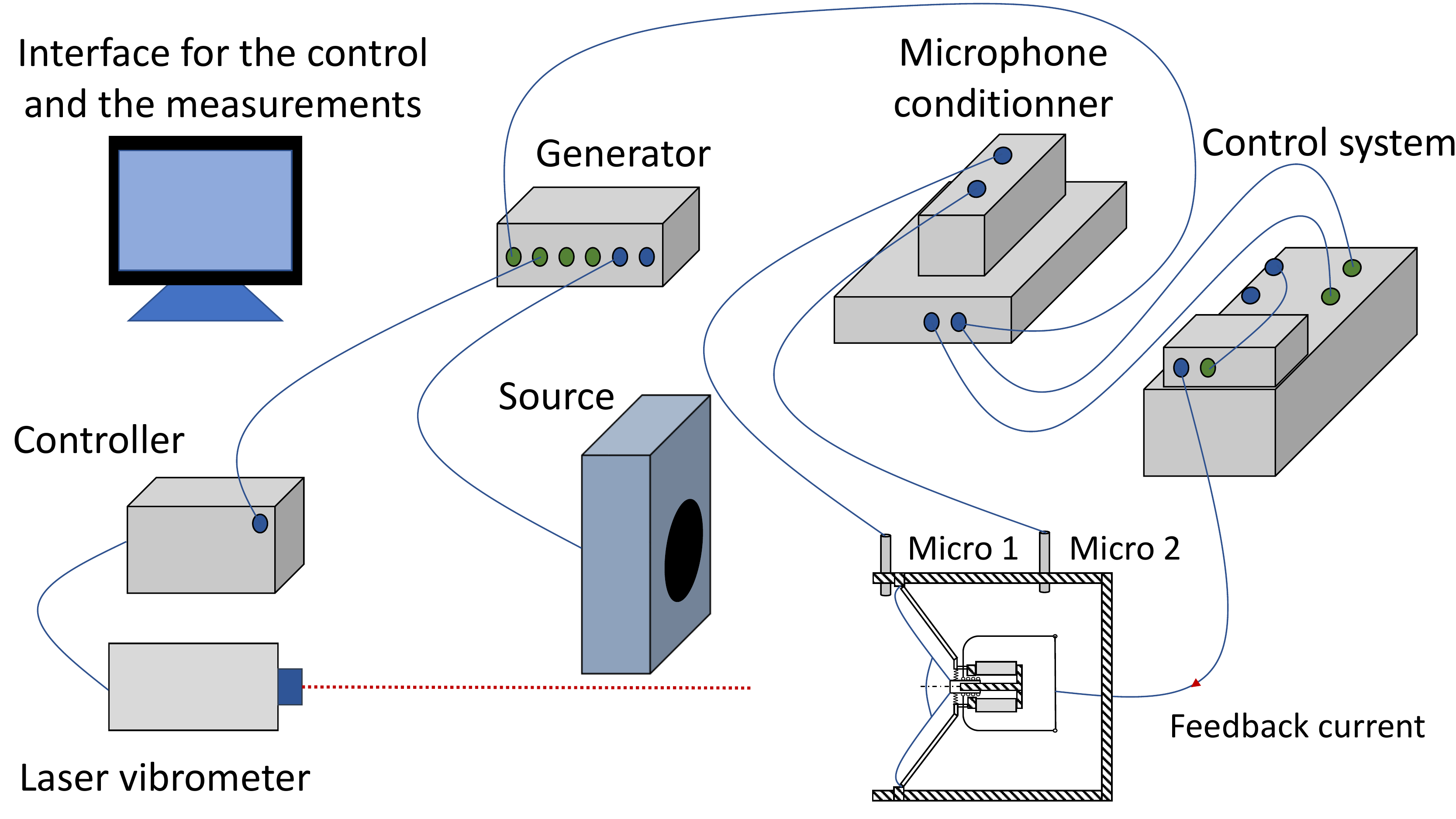}
	\centering
	\caption{\label{fig2} Experimental set up used for applying the feedback current control on the considered closed-box loudspeaker. }
\end{figure}

For implementing the desired controls, two PCB Piezotronics Type 130D20 ICP microphones (nominal sensitivities $G_{mic}=45$ mV/Pa) are employed for sensing respectively the front pressure $p_f$ and the rear pressure $p_b$ of the loudspeaker diaphragm, as illustrated in Fig.~\ref{fig2}. In the case of purely nonlinear (or linear) control, only the measured rear (or front) pressure is used by the control system to generate an output current to the ER, whereas for achieving the hybrid active control, both pressures $p_f$ and $p_b$ are used. The control law is operated through a National Instrument CompactRio FPGA platform, set via LabVIEW 2017 (32bit). The current-drive amplifier feeding back the ER enables the conversion from voltage to current with a gain of $G_{ui}\approx \SI{9.63}{mA/V}$. 

For the acoustic measurements, a Tannoy loudspeaker driven by a signal generator is employed for exciting the whole system. The front pressure is sensed with the same microphone that is used for the control implementation (placed near the front face of the loudspeaker). The membrane axial velocity of the ER is captured by a laser vibrometer focused on the loudspeaker diaphragm. Depending on the definition of the absorption coefficient (accounting the nonlinear effects or not), the absorption performance of the developed AER can be properly characterized through a judicious front pressure and velocity measurement scheme.

Since the assessment of the developed nonlinear AERs needs the comparison with the linear ones, we use both definitions of the absorption coefficient given in Section \ref{Sec:alpha}. For the linear cases, the absorption coefficient $\alpha_L$ is determined from the frequency domain measurements. While a bidirectional sweep sine from \SI{20}{Hz} to \SI{820}{Hz} with sweep rate of \SI{20}{mdec/s} is delivered to the sound source, the transfer function between the front sound pressure $p_f$ and the membrane velocity $v$ is first estimated. $\alpha_L$ is then derived according to Eq.\eqref{eq9}. In the nonlinear cases, the absorption coefficient $\alpha_{NL}$ is determined from time domain measurements, since it requires to extract the pressure amplitudes of all generated harmonics. To this end, a step-wise monochromatic sine excitation with varying frequency within the range of interest $[50Hz, 500Hz]$ is used. For the sake of simplicity, a fine frequency step of $\SI{2}{Hz}$ is employed around the resonance of the ER and a coarser frequency step of $\SI{10}{Hz}$ is employed further away from the resonance. The time signals are acquired and recorded with resolution of $\SI{78}{\mu s}$ and duration of $10$ s. Based on these measurements, the Fourier transform allows amplitude estimation of all harmonic components of the measured quantities. Thus the absorption coefficient $\alpha_{NL}$ can be derived according to Eq.~\eqref{eq10}. However the incident pressure amplitude needs to be estimated in advance over the whole frequency of interest, which is proceeded in the calibration step described in the following.

\begin{figure}
	\includegraphics{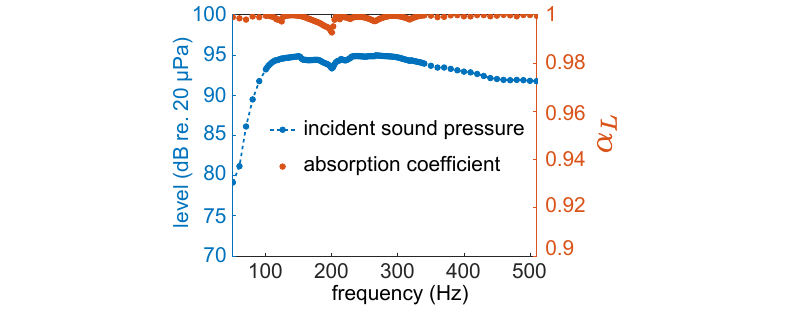}
	\centering
	\caption{\label{fig2_inc} Incident sound pressure level (dB) measured with the microphone nearby the loudspeaker diaphragm when nearly perfect absorption ( $\alpha_L > 0.99$) is achieved with linear AERs over the frequency range of interest $[\SI{50}{Hz}, \SI{500}{Hz}]$.}
\end{figure}

In different studies of nonlinear resonant systems reported so far, strong input intensities are typically required to trigger nonlinearities, such as the NES where the sound pressure levels for activation and observation of nonlinear effects are in the range of 160 dB (1 kPa) \cite{Mattei_2010, Bryk_JSV}. Conversely, in the presented work, we focus on excitation levels that are 3 orders of magnitudes weaker, corresponding to maximum pressure amplitude in the range of $1$ Pa in front of the AER. Consequently, the generated nonlinear effect only results from the proposed active control. Then, in order to calibrate the incident sound pressure delivered to the ER, the sound pressure is measured near the diaphragm with the ER set perfectly absorbent, at different frequencies within the range of interest. In that view, several control laws are applied to the ER, so that it behaves as a narrow-band perfect absorber with various central frequencies, i.e., with $R_{st}=Z_c$ and with different values of $(\mu_1, \mu_2) <1$. These settings allow achieving $\alpha_L > 0.99$ over each segmented frequency range respectively. In this way, an anechoic termination can be provided in an active manner for the whole frequency range of interest $[\SI{50}{Hz}, \SI{500}{Hz}]$. Fig.~\ref{fig2_inc} presents the incident sound pressure levels (dB) measured in front of the diaphragm, at all considered frequencies when the ER is set absorbent, with the sound source located at a fixed distance from the ER. It can be seen from Fig.~\ref{fig2_inc} that the incident pressure presents an amplitude around $1.1$ Pa ($94.8$ dB), especially in the range of $[100 Hz, 400 Hz]$ where the nonlinear effect is strong. These obtained incident pressure levels are exploited in the following as a reference to derive the proportion of energy reflected through higher harmonics ($n\omega$ with $n\geq 2$).

\section{Nonlinear impedance control} \label{NL control}
\begin{figure}
	\includegraphics{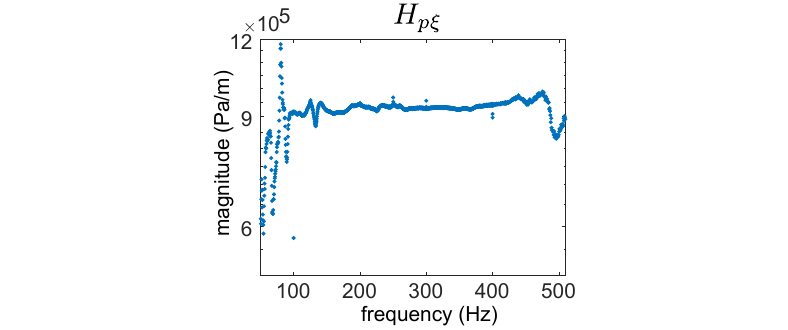}
	\centering
	\caption{\label{fig3} Magnitude of the measured transfer function $H_{p\xi}$ between the rear pressure $p_b$ (Pa) and the loudspeaker diaphragm displacement $\xi$ (m) within the frequency range $[\SI{50}{Hz}, \SI{500}{Hz}]$.}
\end{figure}

A first preliminary test is performed to validate the proportionality between the measured rear pressure $p_b$ and the diaphragm displacement $\xi$, as assumed in Eq.~\eqref{eq2}. For that, the transfer function defined in the frequency domain as $H_{p\xi}=P_b(j\omega)/\Xi(j\omega)=j\omega P_b/V$, where $\Xi(j\omega)$ denotes the frequency response of the displacement $\xi(t)$, is estimated  for frequencies under $\SI{500}{Hz}$, and is displayed in Fig.~\ref{fig3}. The measurement confirms that this transfer function is almost constant in the frequency range of interest, and that the proportionality factor is averaged to $925\times10^{3}$ $\si{Pa/m}$. 

\begin{figure}
	\includegraphics{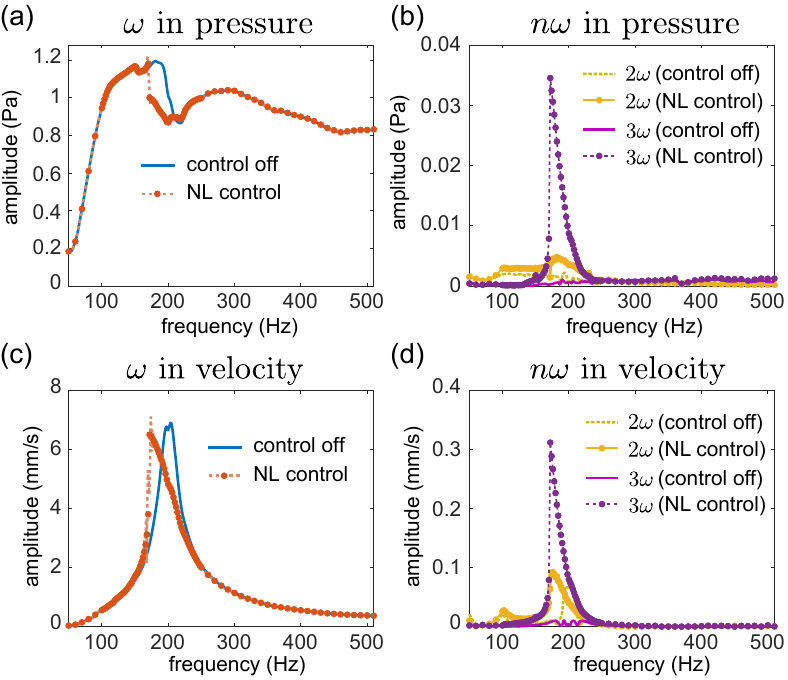}
	\centering
	\caption{\label{fig4} Nonlinear frequency responses of the measured acoustic pressure in front of the AER and of the sensed diaphragm velocity, under control off ($\beta_{NL}=0$) and under pure nonlinear control on ($\beta_{NL}=20$) respectively. Fundamental wave components of the front pressure and the diaphragm velocity are extracted from time domain measurements at each excitation frequency $\omega$ and are shown in (a) and (c) respectively, while the amplitudes of the second and the third harmonic (at frequency $2\omega$ and $3\omega$ respectively) are extracted as well and are presented in (b) for pressure and in (d) for velocity.}
\end{figure}

After validation of the required linear relation, a pure nonlinear control law defined with $i=i_{NL}$ (Eq.~\eqref{eq7}) is applied to the ER. Fig.~\ref{fig4} shows the different experimental results achieved when the control is off ($\beta_{NL}=0$) and when it is on ($\beta_{NL}=20$), respectively. For a better comparison between the two cases, the whole measurements are carried out in the time domain under step-wise sine excitations. Since the sound source is located at fixed position, the same incident excitation amplitude as the one measured during the calibration phase is considered, namely as low as 1.1 Pa in front of the ER, as presented in Fig.~\ref{fig2_inc}. Fig.~\ref{fig4}(a) and Fig.~\ref{fig4}(c) present the corresponding linear frequency responses. The amplitudes of the second and third harmonics are extracted as well for both control cases, and are reported in Fig.~\ref{fig4}(b) and Fig.~\ref{fig4}(d) for the front total pressure and the diaphragm velocity, respectively.

When the control is off, the (linear) resonance of the SDOF ER can be clearly identified at the expected frequency ($\SI{200}{Hz}$) in the linear frequency response of the diaphragm velocity (see Fig.~\ref{fig4}(c)). Although the (ER) loudspeaker is not perfectly linear, the generated higher harmonics stay negligible in the passive case, with maximum pressure amplitude of less than two thousandths of the fundamental wave as can be seen in Fig.~\ref{fig4}(b). The chosen weak incident pressure level ($\approx 1.1 \text{Pa}$) ensures that the amplitude of total pressure remains also in the range of 1 Pa (with maximum value of 1.2 Pa) even around the resonance. When the nonlinear active control is on, a typical nonlinear resonance frequency shift can be observed (see Fig.~\ref{fig4}(c)). The third harmonic component is significantly increased around the nonlinear resonance of the AER owing to the defined cubic nonlinear control law, enabling a normalized maximum pressure amplitude of around $0.035$ with respect to the incident wave which corresponds to an energy proportion nearly $0.13\%$.

Moreover, a slight increase of the second harmonic can also be observed when the active control is on, around the nonlinear resonance, and around $100$ Hz, i.e., at half of the natural resonance frequency. Indeed, an excitation at $100$ Hz enables the frequency match between the second harmonic and the resonance of the ER, favoring the manifestation of $2\omega$ even without external control, as shown in Fig.~\ref{fig4}(b) and Fig.~\ref{fig4}(d). This phenomenon has already been revealed in previous theoretical and numerical works on different sub-wavelength resonators \cite{xinxin_JAP, xinxin_PRE}. However, the third harmonic remains largely dominant around the nonlinear resonance of the AER (see Fig.~\ref{fig4}(b) and Fig.~\ref{fig4}(d)). Therefore, the applied nonlinear active control actually favours the manifestation of a cubic nonlinear resonator, as intended in the specified control law. Otherwise, comparing to the illustrated second and third harmonics, other higher harmonics ($n>3$) are even weaker, with maximum amplitude less than fortieth of that of the third harmonic, thus they are not reported here.

\begin{figure}
	\includegraphics{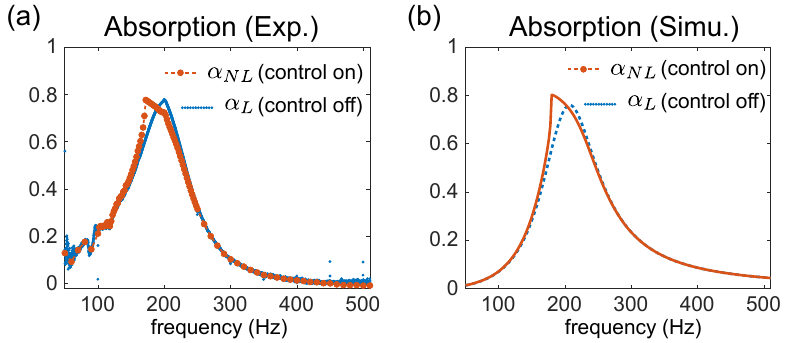}
	\centering
	\caption{\label{fig5} Absorption curves of the achieved nonlinear AER, obtained with the definition of absorption coefficient of Eq.~\eqref{eq10} adapted to nonlinear systems. Same configurations as in Fig.~\ref{fig4} are considered, with nonlinear parameter set as $\beta_{NL}=0$ (blue dotted line) and $\beta_{NL}=20$ (red dash-dotted lines) respectively. Both experimental (Exp.) (a) and simulation (Simu.) (b) results are presented.}
\end{figure}

The focus is hereafter put on the effect of the nonlinear control on the sound absorption properties of the achieved nonlinear AER. From the Fourier analysis of the measured pressure and velocity shown in Fig.~\ref{fig4}, the linear part of the absorption coefficient defined in Eq.~\eqref{eq9}, denoted as $\alpha_{L}$, can be determined by extracting the fundamental components (Fig.~\ref{fig4}(a) and Fig.~\ref{fig4}(c)). Then, the pressure amplitudes of the second and third harmonics of Fig.~\ref{fig4}(b), divided respectively by that of the incident wave presented in Fig.~\ref{fig2_inc}, allows the estimation of the energy proportion reflected through these higher harmonics. Thus following Eq.~\eqref{eq10}, the desired absorption coefficient $\alpha_{NL}$ can be derived by subtracting the above reflected energy part from the linear absorption part $\alpha_L$. For the same cases of control on and off considered in Fig.~\ref{fig4} (identified by $\beta_{NL}=20$ and $\beta_{NL}=0$ respectively), Fig.~\ref{fig5}(a) illustrates the derived absorption coefficient ($\alpha_L$ for passive case, $\alpha_{NL}$ for nonlinear case) under the same weak excitation level as before.

The comparison between such two cases shows that the achieved nonlinear AER allows primarily for broadening the absorption bandwidth towards low frequency in the vicinity of the resonance, corresponding to an increase in the impedance bandwidth (measured as the bandwidth over which the ER impedance magnitude is lower than $\sqrt{2} Z_{min}$, where $Z_{min}$ is the minimum magnitude of the impedance) from around $\SI{23}{Hz}$ to $\SI{35}{Hz}$, namely $50\%$ of increase. Optimal absorption improvement occurs at the nonlinear resonance frequency ($172$ Hz) where $\alpha_{NL}$ increases from $0.61$ to $0.78$. However, according to the frequency responses of all generated harmonics presented in Fig.~\ref{fig4}, only a tiny fraction of energy is transferred into higher harmonics with a maximum proportion of only $0.13\%$. The nonlinear effect introduced by the proposed active control manifests mainly in enhancing sound absorption around the resonance of the AER, while producing only negligible distortion. The absorption curve with definition of $\alpha_{NL}$ including energy radiation of all generated harmonics (e.g., Fig.~\ref{fig5}(a)), in parallel with the frequency responses of the nonlinear components (e.g., Fig.~\ref{fig4}(b)), allows for a complete assessment of the nonlinear effect on the absorption performance of the achieved AER. The combination of these two types of acoustic quantities will be thus considered for illustrating the results of all the following, more advanced, control configurations.

In order to validate that the observed nonlinear behavior results from the defined nonlinear control rule, a numerical simulation based on the classical fourth-order Runge-Kutta (RK4) integration method \cite{Hairer1993} is herein implemented via Matlab. In the simulation, the time delay $\tau$ between the input and the output of the control system is accounted for, since such delay can make the resulting absorption coefficient different from the one obtained directly via Eq.~\eqref{eq3}. Thus, with the defined feedback current $i(t)=G_{ui}\beta_{NL}(p_d(t)G_{mic})^3 \propto \xi^3(t)$, the full problem under consideration is described by the modified motion equation as: 
\begin{align}
M_{ms} \frac{d^2 \xi(t)}{dt^2} =& p_f(t)S_{d}-R_{ms}\frac{d \xi(t)}{d t}-\frac{1}{C_{mc}}\xi(t) \nonumber \\
 &-Bli(t-\tau)H(t-\tau),
\label{eq11}
\end{align}
where $H(t-\tau)$ is the Heaviside function which equals to $1$ for $t \ge \tau$ and to zero for else.

For the sake of accuracy, a step-wise monochromatic source with duration of $\SI{20}{s}$ at each frequency step is considered in the simulations. For each discrete frequency, the absorption coefficient as defined in Eq.~\eqref{eq9} is derived from the total front acoustic pressure and the velocity that are determined by solving numerically the above motion equation Eq.~\eqref{eq10}. Regarding the considered time delay, using a sweep step of $1\times10^{-5}$ \si{s} in simulation, it is found to be $\tau=6\times10^{-5}$ \si{s} by fitting the experimental results. Fig.~\ref{fig5}(b) shows the simulation results of the defined absorption coefficient for both control off ($\beta_{NL}=0$) and pure nonlinear control cases ($\beta_{NL}=20$), under sine excitation performed with a frequency step of $\SI{2}{Hz}$ in the range of $[\SI{50}{Hz}, \SI{500}{Hz}]$.

Although the parameter estimation method employed for extracting the physical parameters ($M_{ms}$, $C_{mc}$, $R_{ms}$ and $Bl$) could be further improved, the comparison between experiments and simulations on the absorption curve, as well as the investigation of higher harmonic generations, is excellent.

Finally, note that in both Fig.~\ref{fig4} and Fig.~\ref{fig5}, the tested nonlinear configuration corresponds to a value $\beta_{NL}=20$, identified as the threshold above which saturation occurs under the considered excitation level. We additionally verified that the same absorption curve can be obtained at lower excitation levels simply by increasing the value of $\beta_{NL}$, which is a clear advantage compared to other passive nonlinear systems reported in the literature. While the bandwidth increase towards low frequency is considerable, the performed nonlinear control can be further controlled and improved by combining with a linear one. Section \ref{hybrid control} will show how such hybrid control allows improving the absorption performance of the ER compared to the case of pure nonlinear control.

\section{Combination of linear and nonlinear impedance control laws} \label{hybrid control}
Different linear active control laws are taken into account in this section and combined with the previously presented nonlinear control. This allows modifying the whole dynamics of the ER. More specifically, the resonance frequency (at which the absorption coefficient is maximal, or the impedance is minimal and purely resistive) can be tuned through the linear control parameter ratio $\mu_2/\mu_1$. Moreover, it has been demonstrated that the absorption bandwidth depends primarily on the amount $S_d/M_{as}$ \cite{etienne2016}. Accordingly, this section considers the variation of design parameter $\mu_2$ assigned to the compliance, while choosing mass factor $\mu_1=1$ so that the bandwidth of absorption of the ER remains nearly unchanged in the linear regime. A brief discussion about the mass factor is given at the end of this section.

\begin{figure}[htbp]
	\includegraphics{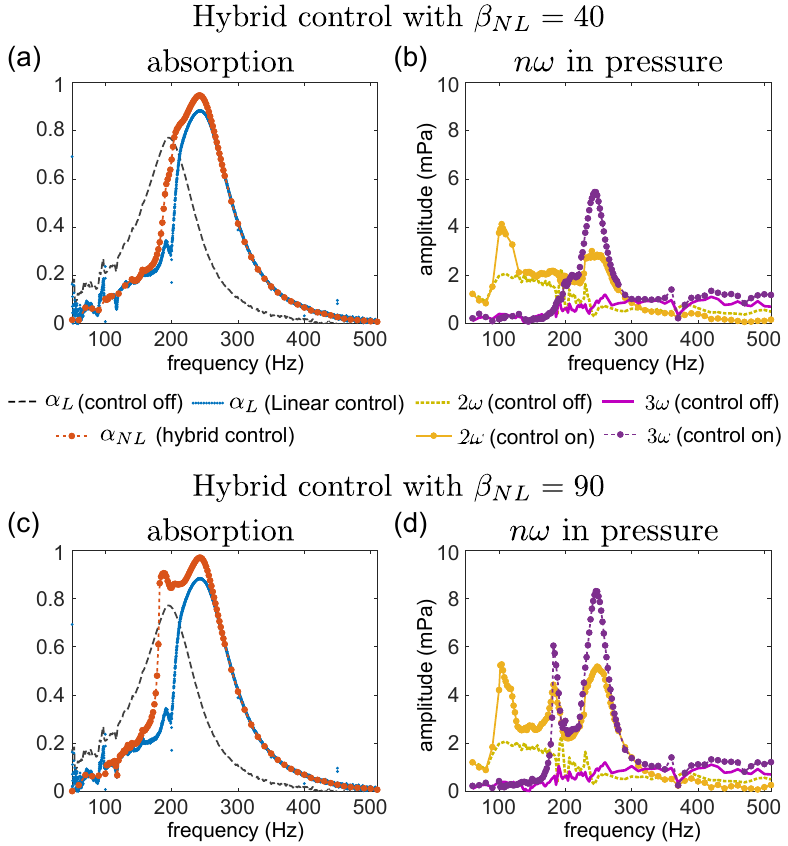}
	\centering
	\caption{\label{fig6} Absorption curves (with $\alpha_{NL}$ defined by Eq.~\eqref{eq10}) of the achieved nonlinear AER (red dash-dotted lines) under a hybrid control with $\mu_1=1$, $\mu_2=1.5$, $R_{st}=0.5Z_c$ and $\beta_{NL}=40$ (a) or $\beta_{NL}=90$ (c) respectively. Absorption results achieved with both cases of control off (black dashed lines) and of pure linear control (blue dotted lines) are shown as well for comparison. For a better demonstration, pressure amplitudes of the generated second and third harmonics are illustrated in (b) and (d) for the two considered hybrid control cases in parallel with absorption curves respectively. Sine excitations are performed with fixed level to deliver the same incident pressures in front of the AER as presented in Fig.~\ref{fig2_inc}. }
\end{figure}

First, a linear control law with $\mu_1=1$ and $\mu_2=1.5$ is considered. It maintains the original absorption bandwidth of the ER but shifts the maximum absorption slightly from $\SI{200}{Hz}$ to $\SI{240}{Hz}$. Regarding the target resistance $R_{st}$, a total absorption ($\alpha=1$) is achievable at the target frequency $f_{st}$ when $R_{st}$ coincides with the specific acoustic impedance of air $Z_c$. With such linear configuration, the nonlinearity provided by the hybrid control can only enable the enlargement of the absorption bandwidth. Hence, with a view to assessing the overall effect of the resulting nonlinearity, we first assign a target resistance different from $Z_c$ for the present linear control law (defined by $\mu_1=1$ and $\mu_2=1.5$), i.e., $R_{st}=0.5 Z_c$.

The desired hybrid control is identified by the feedback current being $i(t)=i_L(t)+i_{NL}(t)$ with linear part $i_L(t)$ satisfying the aforementioned target impedance, and with nonlinear part $i_{NL}(t)$ obtained by implementing the cubic product of the rear pressure (Eq.~\eqref{eq7}). Fig.~\ref{fig6} shows the achieved absorption curve and the corresponding frequency dependence of the generated second and third harmonics, for nonlinear configurations defined by $\beta_{NL}=40$ (Fig.~\ref{fig6}(a) and Fig.~\ref{fig6}(b)) and $\beta_{NL}=90$ (Fig.~\ref{fig6}(c) and Fig.~\ref{fig6}(d)) respectively. The control off case ($i=0$) and the pure linear control case ($i_{NL}=0$) are illustrated as well in both Fig.~\ref{fig6}(a) and Fig.~\ref{fig6}(c), with black dashed lines and blue dotted lines, respectively.

In comparison with the pure nonlinear control presented in Section \ref{NL control}, the hybrid control allows for further improvement of the sound absorption performance. With the presented linear part of control, the nonlinear parameter can be even increased and exceed $\beta_{NL}=90$ without saturation, thus enabling a bandwidth of efficient absorption ($\alpha>0.8$ as explained in \cite{etienne_IEEE}) of around $\SI{80}{Hz}$, while increasing the absorption magnitude with maximum value up to 0.98. Moreover, when compared to the pure linear control case that allows the efficient absorption bandwidth of $\SI{40}{Hz}$ and a maximum absorption value of around $0.89$ (blue dotted lines in Fig.~\ref{fig6}), the nonlinear part of the proposed hybrid control with nonlinear parameter set as $\beta_{NL}=90$, is capable of significantly improving absorption performance, i.e., doubling the efficient absorption bandwidth and yielding a nearly perfect absorption near the target (nonlinear) resonance frequency.

Nevertheless, regarding the third harmonic, although it can be amplified by increasing the value of the nonlinear parameter, the present hybrid control limits the third harmonic generation to an amplitude less than a quarter of that presented with the pure nonlinear control, as evident when comparing Fig.~\ref{fig6}(d) and Fig.~\ref{fig6}(b) to Fig.~\ref{fig5}(b). Under such hybrid control, two maximums are visible in the frequency response of the third harmonic $3\omega$. The most important one, at frequency of around $\SI{248}{Hz}$, corresponds to the targeted resonance which is prescribed by the linear control law and slightly shifted due to the nonlinear effect, and the other one appears in the vicinity of the natural resonance and is linked to the mismatch in the mechanical parameter estimation used for the resonance adjustment through linear control laws. 

Since the ER is never perfectly linear, a second harmonic component is also present, even in the passive case. When the hybrid control is applied, in addition to the two maxima occurring at the same excitation frequencies as the third harmonic owing to the presence of the resonance, a third one can also be noticed at frequencies ranging from half of the natural resonance up to half of the target resonance of the AER. The triggered nonlinear effect favors the second harmonic generation around this range, because of the correspondence between the generated second harmonic and the ER resonances. Conversely, the third harmonic can not be triggered around $100$ Hz, since the applied nonlinear control law can not play an important role when far from the resonances. Still, when close to the aforementioned two resonance frequencies, the third harmonic prevails over the second harmonic due to the cubic nonlinearity introduced through active control.

\begin{table}[htbp]
\caption{\label{table_inc} The required value of nonlinear parameter $\beta_{NL}$ that leads to the same nonlinear absorption curve as that of Fig.~\ref{fig6}(c) under different incident pressure levels in front of the AER achieved by varying excitation levels.}
\begin{center}
\begin{ruledtabular}
\begin{tabular}{lccccc}
\textbf{Incident pressure level} (dB) & \multicolumn{1}{c}{88.8} & \multicolumn{1}{c}{91.3} & \multicolumn{1}{c}{94.8} & \multicolumn{1}{c}{97.5} & \multicolumn{1}{c}{99.6} \\
\hline
\rule{0pt}{12pt}\textbf{Required value of $\beta_{NL}$} &  \multicolumn{1}{c}{400} &  \multicolumn{1}{c}{180} &  \multicolumn{1}{c}{90} &  \multicolumn{1}{c}{50} &  \multicolumn{1}{c}{30}\\
\end{tabular}
\end{ruledtabular}
\end{center}
\end{table}

Additionally, the influence of the excitation level is also studied. Table \ref{table_inc} shows the required value of nonlinear parameter $\beta_{NL}$ that leads to the same absorption curve as that of Fig.~\ref{fig6}(c), as a function of the incident pressure level. One can notice that, as the input intensity is decreased, the absorption performance can still be enhanced by increasing the value of the nonlinear parameter. The ability of such hybrid control in improving sound absorption at low intensities is herein confirmed, regardless of excitation levels.

\begin{figure}[htbp]
	\includegraphics{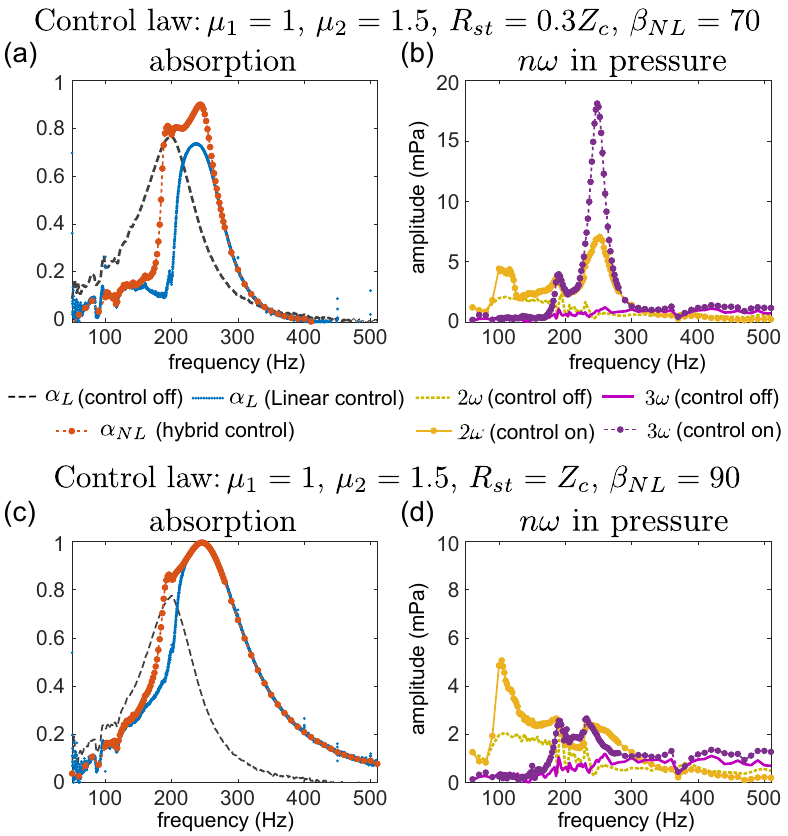}
	\centering
	\caption{\label{fig7} Absorption curves of the achieved nonlinear AER under different hybrid controls with linear design parameters defined as $\mu_1=1$, $\mu_2=1.5$, and with linear target resistance being $R_{st}=0.3Z_c$ (a) and  $R_{st}=Z_c$ (c) respectively. Nonlinear part of control is applied with achievable value of nonlinear parameter set as $\beta_{NL}=70$ and $\beta_{NL}=90$ respectively. For both control configurations, higher harmonic generation is taken into account in the definition of absorption coefficient $\alpha_{Nl}$ as described by Eq.~\eqref{eq10}, pressure amplitudes of second and third harmonics are illustrated in (b) and (d) in parallel with the corresponding absorption curves (a) and (c) for the two control cases respectively. Pure linear control results with $\beta_{NL}=0$ (black dashed line) and the control off case (blue dotted lines) are also presented for the comparison with hybrid control result.}
\end{figure}

Following the previous configuration, Fig.~\ref{fig7} presents the absorption curve achieved by modifying the linear part of the control law. Here, the reactive parameters are set so that to preserve the same linear target resonance frequency $f_{st}$ as in Fig.~\ref{fig6} ($\mu_1=1$ and $\mu_2=1.5$), while varying the target resistance to $R_{st}=0.3Z_c$ (Fig.~\ref{fig7}(a) and Fig.~\ref{fig7}(b)) and to $R_{st}=Z_c$ (Fig.~\ref{fig7}(c) and Fig.~\ref{fig7}(d)) respectively. The nonlinear parameter is set as high as possible below the saturation threshold. According to the comparison between the three control cases presented in Fig.~\ref{fig6} and Fig.~\ref{fig7}, the nonlinear effect is more pronounced as the target resistance $R_{st}$ decreases. The amplitude of the third harmonic generated by the control with $R_{st}=0.3Z_c$ shows a maximum higher than twice of that achieved with $R_{st}=0.5Z_c$ or $R_{st}=Z_c$, while a lower nonlinear parameter ($\beta_{NL}=70$ instead of $90$) is provided in such control. Moreover, a rather low resistance allows not only broadening the absorption bandwidth, but also increasing the absorption level. Indeed, as can be seen in Fig.~\ref{fig7}(a), the linear control with $R_{st}=0.3Z_c$ allows a maximum magnitude of absorption coefficient of around $0.73$, whereas it can exceed $0.9$ with hybrid control, enabling an effective absorption bandwidth ($\alpha > 0.8$) of around $\SI{62}{Hz}$.

However, when the target resistance is equal to the specific acoustic impedance of air ($R_{st}=Z_c$), although the cubic nonlinearity is less triggered (identified by weak generation of third harmonic), the final absorption result appears to be optimal. With such linear control law that provides a nearly total absorption at the targeted resonance frequency, the nonlinear part of the corresponding hybrid control leads dominantly to an increase of the absorption bandwidth. The bandwidth over which effective sound absorption ($\alpha > 0.8$) is achieved has been extended from $\SI{75}{Hz}$ through pure linear control to $\SI{95}{Hz}$ through hybrid control, as shown in Fig.~\ref{fig7}(c).

\begin{figure}[htbp]
	\includegraphics{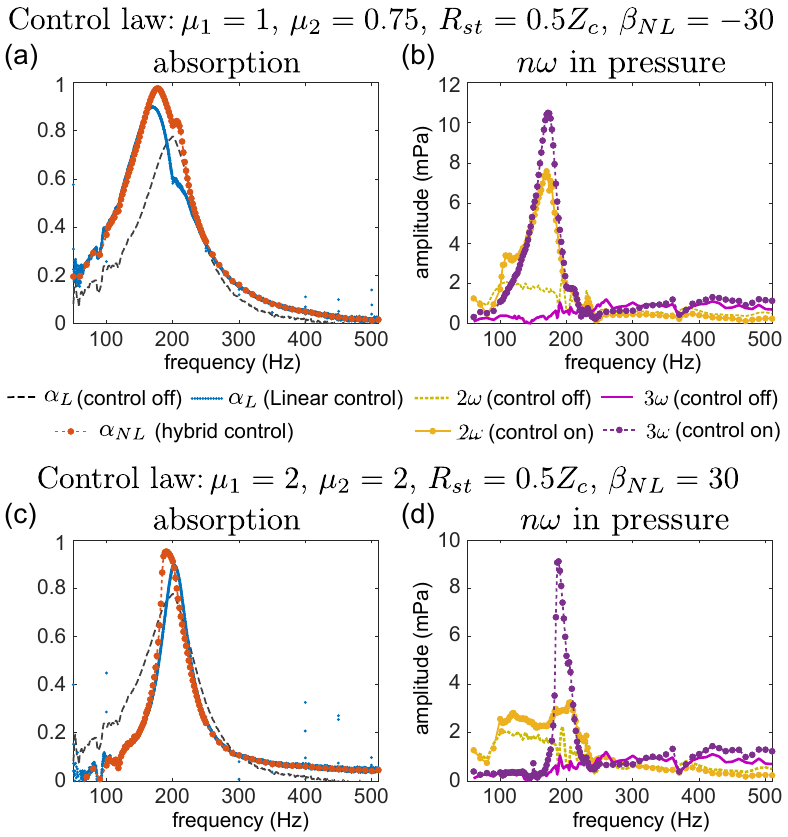}
	\centering
	\caption{\label{fig8} Absorption curves of the achieved nonlinear AER and the associated pressure amplitudes of the second and third harmonics, by considering the definition of absorption coefficient suitable for nonlinear systems (Eq.~\eqref{eq10}). Two hybrid control results are presented, identified by the linear part defined by $\mu_1=1$, $\mu_2=0.75$ and $R_{st}=0.5Z_c$ (a) and (b) and $\mu_1=\mu_2=2$ and $R_{st}=0.5Z_c$ (c) and (d), respectively. Nonlinear part of control is applied with achievable value of nonlinear parameter $\beta_{NL}$ being $\beta_{NL}=-30$ ((a) and (b)) and $\beta_{NL}=30$ ((c) and (d)) respectively. Absorption results of pure linear control with $\beta_{NL}=0$ (blue dotted lines) and the control off case (black dashed lines) are also presented for both configurations.}
\end{figure}

Similar to the configurations presented in Fig.~\ref{fig6} and Fig.~\ref{fig7}, Fig.~\ref{fig8} shows the absorption curves obtained with different reactive parameters of the linear control part ($\mu_1$ and $\mu_2$), while maintaining the target resistance to $R_{st}=0.5 Z_c$ and the nonlinear control law with parameter $\beta_{NL}$ as large as possible, provided that the whole system remains stable (without saturation).

A linear configuration allowing shifting the (linear) resonance towards low frequency is firstly considered and presented in Fig.~\ref{fig8}(a) and Fig.~\ref{fig8}(b), with linear parameters set as $\mu_1=1$, $\mu_2=0.75$ targeting a resonance at around $176$ Hz. When the resonance frequency of the AER is linearly tuned below the natural resonance frequency of the passive ER (softening instead of stiffening), the nonlinear parameter $\beta_{NL}$ needs to be negative to enable absorption improvement. With the design parameter $\mu_2$ decreasing until $0.5$, an absolute value of $30$ for $\beta_{NL}$ is proved consistently achievable in measurement without saturation. Comparing to the cases of $\mu_1 < \mu_2$ (see Fig.~\ref{fig6} and Fig.~\ref{fig7}), in the present configuration, the nonlinear component of the hybrid control leads to the same trend in absorption improvement, i.e., bandwidth enlargement and amplitude increase mainly within the frequency range bounded by the target resonance and the natural resonance of the ER. As a result, even though the current control appears less advantageous than the previous ones, a maximum absorption up to $0.98$ can still be achieved together with the efficient absorption bandwidth extended from $36$ Hz through pure linear control to $60$ Hz through hybrid control.

In addition to the previous configuration, Fig.~\ref{fig8}(c) and Fig.~\ref{fig8}(d) present the control results for linear target impedance defined with $\mu_1=\mu_2=2$ preserving the (linear) resonance frequency of the passive ER but enabling a higher quality factor (or in other words, a narrower bandwidth of absorption). The target resistance is set to $R_{st}=0.5Z_c$ as in the previous case of Fig.~\ref{fig8}(a). With $\mu_1=\mu_2 > 1$, the hybrid control leads to a similar absorption result as the pure nonlinear control configuration where $\mu_1=\mu_2 =1$ (see Fig.~\ref{fig4} and Fig.~\ref{fig5}). The triggered nonlinear effect manifests as a slight enlargement of absorption bandwidth towards the low frequency range, enabling the efficient absorption bandwidth changed from $[\SI{194}{Hz}, \SI{212}{Hz}]$ to $[\SI{185}{Hz}, \SI{206}{Hz}]$, and along with an increase of the maximum absorption magnitude from $0.9$ to $0.95$. Although it is also possible to broaden the absorption bandwidth with linear control by defining $\mu_1=\mu_2<1$  \cite{etienne_IEEE, etienne2016}, when such a scheme is implemented in a hybrid control, saturation prevents the nonlinear parameter $\beta_{NL}$ to be increased to the same level as for the cases $\mu_1=\mu_2 \ge 1$. Thus, in such configuration, the generated weak nonlinear effect leads only to a tiny improvement of absorption.

In both configurations presented in Fig.~\ref{fig8}, the generated second and third harmonics always present a maximum amplitude at the shifted target resonance frequency, as in all the previous configurations presented in Fig.~\ref{fig5}, Fig.~\ref{fig6} and Fig.~\ref{fig7}. Under the defined cubic nonlinear control law, the third harmonic remains more important than the second one in the frequency range where the nonlinear effect acts on the absorption performance. Conversely, around $100$ Hz, the second harmonic is more prominent since its frequency coincides with either the natural or the target resonance. However, in all considered control configurations, the generated higher harmonics are consistently very weak compared to the fundamental component over the whole frequency range of interest. The maximum pressure amplitude appears at the third harmonic, of $ 0.035 $ Pa, corresponding to a reflected energy proportion of only $0.13\%$.

Hence, according to the results obtained in this section with different hybrid control laws, we conclude that the nonlinear effect enabled via the active control allows for significantly improving the absorption performance of the ER, while producing only negligible sound wave distortion. Depending on the linear part of the control law, the generated nonlinearity can play a role of variable importance, i.e., either in expanding the bandwidth or simultaneously increasing the magnitude and enlarging the bandwidth of effective absorption. The optimal hybrid control law includes a linear part that slightly shifts the linear maximum absorption and a nonlinear part defined with nonlinear parameter as high as possible provided that no saturation occurs. When compared to the only linear or nonlinear active control, the hybrid control presents more advantages in improving the absorption performance of the achieved AER, thus having the potential to be widely used for future low-frequency sound absorption. 

\section{Conclusion}
Based on an experimental prototype developed for achieving linear active impedance control on a closed-box electrodynamic loudspeaker, a nonlinear active impedance control has been introduced and implemented in the present work. Thanks to the proportionality between the displacement of the loudspeaker diaphragm and the rear pressure, within the low frequency range of interest ($\SI{50}{Hz}, \SI{500}{Hz}$), a nonlinear AER with cubic nonlinearity has been experimentally achieved, allowing its combination with the already existing linear active ER scheme. 

Our study has focused on the absorption performance of the resonator, by first considering a pure nonlinear control, and then a hybrid control that combines linear and nonlinear control laws. With a view to fully analyzing the triggered nonlinear effect, an absorption coefficient accounting for the likely generation of higher harmonics has been defined. Unlike the other nonlinear mechanisms that require significantly high pressure levels to enable the nonlinear effect manifestation, such as reported in the literature on NES used also for the absorption enhancement, the present control architectures allow for sound absorption improvement at much lower excitation levels, while producing negligible distortion. Compared to the employed passive SDOF ER presenting a maximum absorption coefficient of about $0.77$ at its natural resonance frequency, under an incident sound pressure level of around $94.8$ dB (around 1.1 Pa) in front of the resonator, a considerable increase in absorption coefficient above $0.8$ can be achieved through the proposed hybrid control within a frequency range larger than $\SI{80}{Hz}$, while along with only $0.13\%$ of energy reflected through higher harmonics.

In the present work, a cubic nonlinear control law on the diaphragm displacement is taken into account. In order to ensure that the performed control operates as defined, a time domain integration method is used to simulate the full problem. A relatively good agreement has been found between the experimental results and the simulation implementations. Such nonlinear control law is also presented as an active manner to achieve a cubic nonlinear stiffness on the resonator. Additionally, the proposed nonlinear active control not only facilitates the generation of nonlinearities on the ER, but also allows them to be adjustable and reprogrammable which is very difficult to obtain using mechanical nonlinearities.

Nevertheless, since the reported nonlinear and hybrid control results strongly depend on the passive acoustical parameters of the considered ER, i.e., mass $M_{as}$, compliance $C_{ac}$, resistance $R_{as}$ and the force factor $Bl$, that are numerically extracted from two impedance measurements with different electric loads, performance could be further improved by additional measurement, for instance by evaluating the effective area of diaphragm $S_d$. Alternatively, the hybrid control law is investigated herein, i.e., with a linear part restricting the ER to be single-degree-of-freedom and with a nonlinear part focusing on the cubic displacement nonlinearity. In the future, other types of nonlinearity could be achieved through the proposed experimental prototype, and be combined with active linear multiple-degrees-of-freedom ER, with the aim of further improving the sound absorption.

As a perspective, such active control scheme could be employed in the design of acoustic metamaterials, in a view to achieving non-trivial wave phenomena. Indeed, an unit-cell implementing the reported active control scheme, with two microphones (one sensing the front pressure and another the rear pressure related to the diaphragm displacement), could intrinsically present negative effective bulk modulus for instance. Combining a nonlinear law with such a linear active control, a new family of nonlinear active metamaterials with potentially larger bandwidth or multistable functionalities could be developed.

\begin{acknowledgments}
The authors wish to thank Dr. Sami Karkar for his precious advice on the implementation of non-linear stiffness.
\end{acknowledgments}


%
\end{document}